# Morphological diversity of spiral galaxies originating in the cold gas inflow from cosmic webs

Masafumi Noguchi[1*]

[1]Astronomical Institute, Tohoku University, Aoba-ku, Sendai 980-8578, Japan

**Spiral galaxies comprise three major structural components; thin discs, thick discs, and central bulges[1,2,3]. Relative dominance of these components is known to correlate with the total mass of the galaxy[4,5], and produces a remarkable morphological variety of spiral galaxies. Although there are many formation scenarios regarding individual components[6,7], no unified theory exists which explains this systematic variation. The cold-flow hypothesis[8,9,] predicts that galaxies grow by accretion of cold gas from cosmic webs (cold accretion) when their mass is below a certain threshold, whereas in the high-mass regime the gas that entered the dark matter halo is first heated by shock waves to high temperatures and then accretes to the forming galaxy as it cools emitting radiation (cooling flow). This hypothesis also predicts that massive galaxies at high redshifts have a hybrid accretion structure in which filaments of cold inflowing gas penetrate surrounding hot gas. In the case of Milky Way, the previous study[10] suggested that the cold accretion created its thick disc in early times and the cooling flow formed the thin disc in later epochs. Here we report that extending this idea to galaxies with various masses and associating the hybrid accretion with the formation of bulges reproduces the observed mass-dependent structures of spiral galaxies: namely, more massive galaxies have lower thick disc mass fractions and higher bulge mass fractions[4,5,11]. The proposed scenario predicts that thick discs are older in age and poorer in iron than thin discs, the trend observed in the Milky Way (MW) [12,13] and other spiral galaxies[14].**

Complex structure of spiral galaxies poses a formidable challenge for current galaxy formation theory. Galaxy formation process is greatly controlled by how the primordial gas flows into dark matter halos of galaxies and accretes finally onto their central part where stars form. The longstanding paradigm , the shock-heating theory[15], argues that the gas that entered the dark matter halo is heated by a shock wave to a high temperature. Then, it cools by emitting radiation and gradually accretes to the central part ('cooling flow' or 'hot-mode accretion'). This picture predicts a smooth and continuous gas accretion history. Therefore it is not easy to understand why and how the different components of disc galaxies formed, and external origin is invoked in some cases. For example, bulges are sometimes regarded as smaller galaxies swallowed in galaxy mergers[16].

The shock-heating theory was later modified by the 'cold accretion' scenario which claims that shock waves do not arise in some cases and a significant part of the primordial gas stays cold, reaching the galaxy in narrow streams almost in a freefall fashion ('cold accretion')[8,9]. This hypothesis predicts three different regimes for the thermal state and accretion process of the halo gas depending on the total galaxy mass (i.e., the virial mass) and redshift[9,17] (Fig.1). Shock waves do not appear and the cold accretion dominates when $M_{\text{vir}} < M_{\text{shock}}$ (Domain A). Halo gas is totally shock-heated to high temperature and the cooling flow of hot gas dominates in Domain B ($M_{\text{vir}} > M_{\text{shock}}$ and $M_{\text{vir}} > M_{\text{steam}}$). Domain C, for which $M_{\text{shock}} < M_{\text{vir}} < M_{\text{stream}}$, has a unique hybrid structure. Narrow streams of cold gas penetrate ambient hot gas heated by shock waves[9,17].

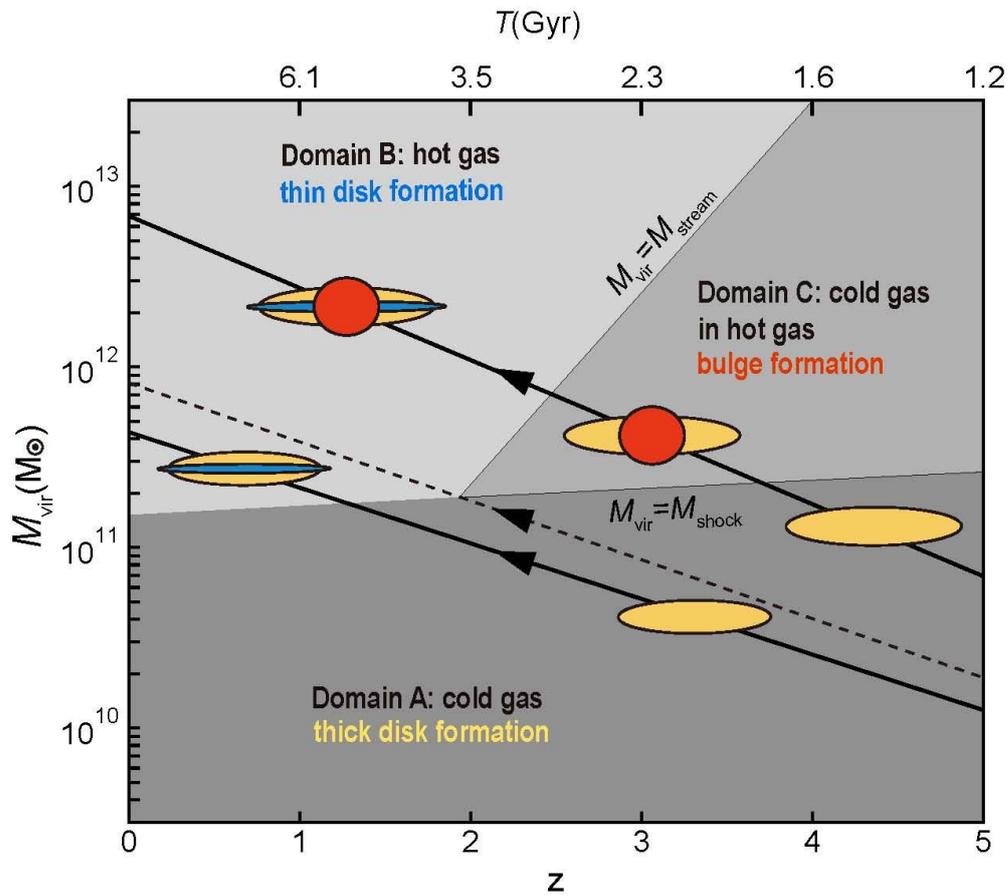

**Figure 1. Three different states of the halo gas according to the cold-accretion theory.** $M_{\mathrm{vir}}$, z, and $T$ denote the virial mass of the galaxy, redshift, and cosmic time, respectively. Formation of thin disc (blue), thick disc (orange), and bulge (red) is indicated schematically.

The previous study[10] proposed that the MW thick disc was formed by the cold accretion and resulting fast star formation in early times, whereas its thin disc was formed later by slow star formation fueled by the hot-mode accretion. This picture explains why the MW thick disc is older in age, poorer in iron, but has relatively more α-elements (O,Mg,Si, etc.) with respect to iron, than the thin disc[12,13].

Accretion diagram in Fig.1 opens an interesting possibility to expand and generalize this idea by applying it to other spiral galaxies with various masses. Two solid diagonal lines in Fig.1 indicate the evolution path of the least and most massive galaxies discussed here, while the dashed line indicates that for the galaxy having the critical mass. We calculate the time evolution of gas accretion rate along this track and examine how different components build up with time by assuming that the rate of star formation is proportional to the accretion rate at each time. In accordance with the previous study[10], the gas accreted in unheated state (Domain A) is deposited in the thick disc, whereas the gas supplied by hot-mode accretion (Domain B) is used to grow the thin disc. The fate of the gas gathered by the halo residing in Domain C is more elusive. By the reason discussed later, we associate it with the bulge component.

As shown in Fig.2**a**, the galaxies with total mass smaller than the critical mass first grow thick discs by the cold-mode gas accretion. After crossing the shock-heating border, gas accretion is switched to hot-mode (Fig.1), and thin discs are formed. Fig.2**c** shows that the mass fraction of thick disc decreases monotonically. The galaxies more massive than the critical mass evolve in a more complicated way (Fig.2**b**). After thick disc formation in Domain A, they enter the hybrid Domain C and develop bulges (Fig.1). Finally thin discs start to form when they leave this domain.

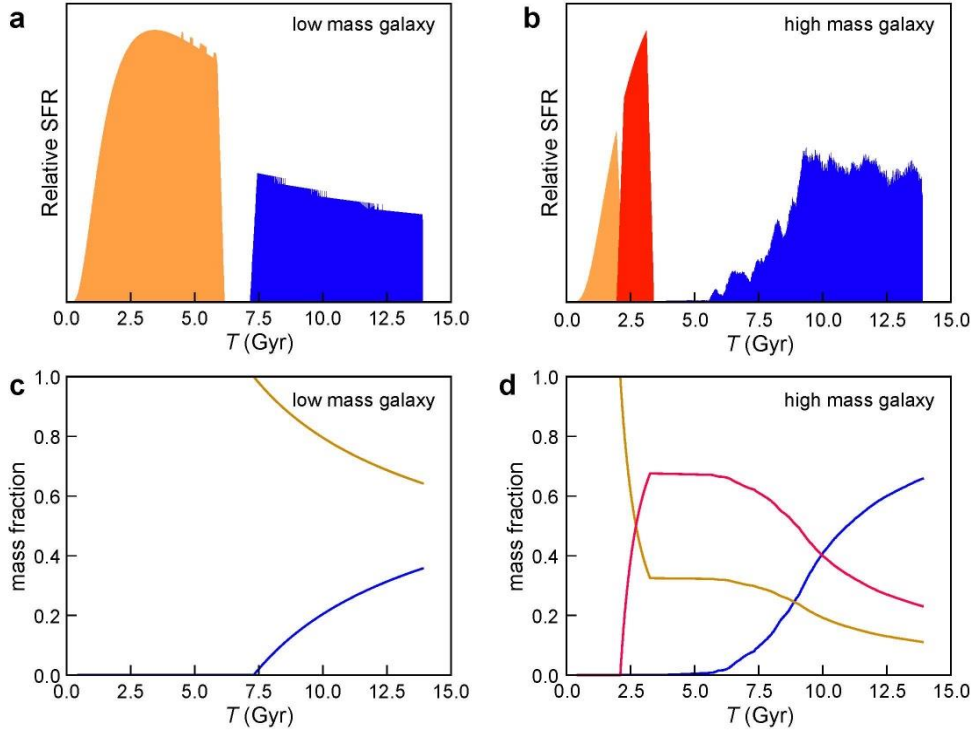

**Figure 2. Star formation history in low-mass and high-mass galaxies.**
**a,b**, Star formation rate (SFR) in the least massive (**a**, with the present virial mass, $M_{\text{vir},0} = 4.3 \times 10^{11} M_\odot$ ) and the most massive (**b**, with $M_{\text{vir},0} = 7.0 \times 10^{12} M_\odot$ ) galaxies. Blue, orange, and red indicate SFR for thin disc, thick disc, and bulge, respectively. SFR is averaged over 20 successive time steps (0.28 Gyr) to smooth out short timescale fluctuations arising in model calculation, and given in arbitrary units. **c,d**, Time variation of the mass fraction relative to the total stellar mass of thin disc (blue), thick disc (orange), and bulge (red).

Galactic bulges in this scenario are not the first mass component to emerge in high-mass galaxies but thick discs predate bulges. Therefore, the bulge mass fraction is initially almost zero and then starts to increase as the bulge develops. In later phase ($T > 7$ Gyr), it decreases because thin disc starts to form (Figure 2**d**). Recent surveys[18,19] suggest that the galaxies in early times are mostly disc-dominated and the significant population of bulge-dominated galaxies emerge relatively recently ($z < 2$ or $T > 3$ Gyr), hinting at the bulge formation after thick disc formation.

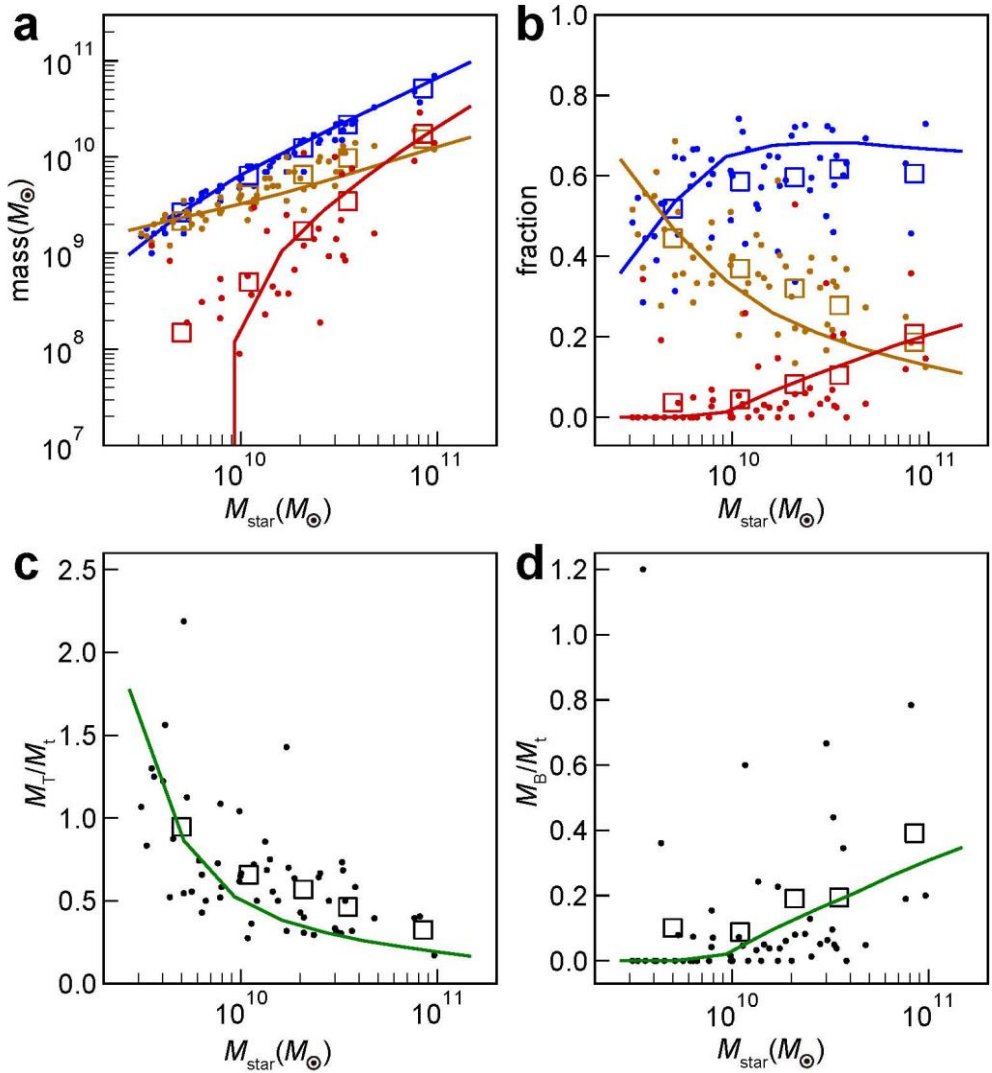

**Figure 3. Masses of various components and their ratios plotted against the total stellar mass of the galaxy.**

The total stellar mass, $M_{star}$, is the sum of the masses of thin disk ($M_t$), thick disk ($M_T$), and bulge ($M_B$). Solid lines show the model. Observed values[5] are indicated by dots, with squares showing the mean in each mass-bin. **a**, Mass of each component (thin disc :blue, thick disc: orange, bulge: red). **b**, The same as **a**, but for the mass fraction of each component. **c, d**, Mass ratios of indicated components.

Figure 3 demonstrates that the model quantitatively reproduces the masses of various components as a function of the total galaxy (stellar) mass for the present-day spiral galaxies[5], although the observation shows a large scatter. Note that the model describes the mean behavior and does not take into account dispersions in the dark matter halo properties at the same virial mass. Both in the model and observation, low-mass galaxies ($M_{star} < 10^{10} M_\odot$) lack bulges (or have only small bulges), with thick discs occupying comparable masses with thin discs (Fig.3**a,b**). Galaxies with larger masses contain increasingly more massive bulge components but the contribution of thick discs decreases. The thin disc is always the most dominant component for the studied mass range. The mass ratio of thick and thin discs decreases monotonically with the total stellar mass (Fig.3**c**), while the bugle-to-thin disc ratio increases (Fig.3**d**). These results are consistent with the behavior of star formation rate in Figure 2 showing that massive galaxies experience more intense hot-mode accretion and an additional hybrid accretion, contributing to thin discs and bulges, respectively.

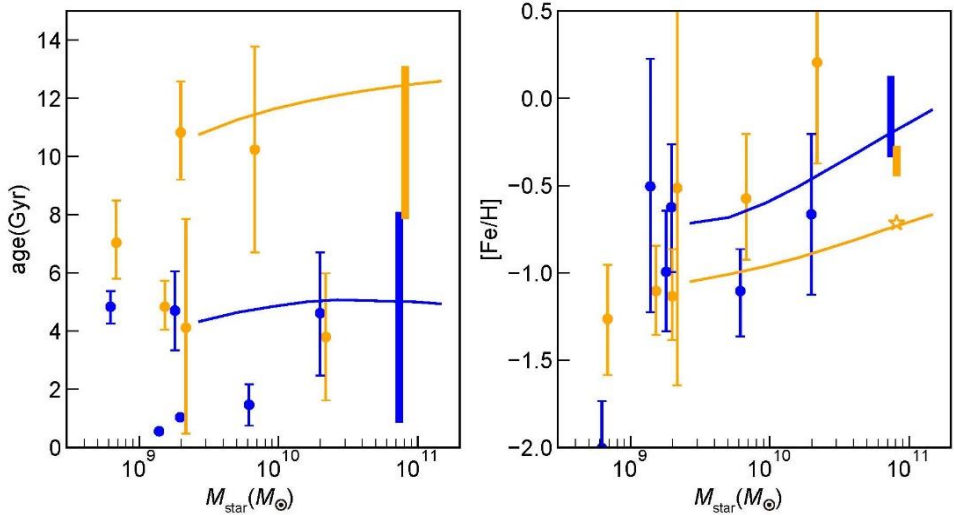

**Figure 4. Mean age and iron abundance of the thick and thin discs plotted against the total stellar mass of the galaxy.**
Blue and orange indicate thin and thick discs, respectively. Solid curves indicate the model. Dots with error bars indicate the observations[14] for external galaxies (Thick disc values are slightly displaced horizontally for clarity). **a**, The mean stellar age. The vertical thick segments indicate the age range estimated for the solar neighbourhood[12]. **b**, The iron abundance ratio. The vertical thick segments indicate the solar neighbourhood observation[13]. The star symbol is the observation for the solar neighborhood thick disc[20].

The two kinds of discs are different also in age and chemical compositions. Fig.4**a** shows that the model thick discs are old components with their stars formed around 10 Gyr ago, whereas thin discs consist of much younger stars, in agreement with the observations for the solar neighbourhood[12] and for other galaxies[14], though the latter cases suffer from larger uncertainties. The chronological order of thick and thin disc formation is also reflected in their chemical properties. The model thin discs include more iron relative to hydrogen than the thick discs (Fig.4**b**) because they are formed from the interstellar gas polluted by iron produced and released by early star formation which made thick discs. Solar neighbourhood shows a similar trend[13,20], though the data for other galaxies show large scatters and no clear trend[14].

We proposed here that thick discs were formed by the cold accretion. Another scenarios propose that they formed by stirring up of the existing thin discs by mergers with smaller galaxies[21] or by the radial migration of kinematically hot stars from inner galactic regions to the solar neighbourhood[22]. It seems, however, difficult to explain the chemical bimodality of MW disc stars[12,13] in these scenarios[22]. Although several processes may contribute to thick disc formation actually, the successful reproduction of structural variations (Fig.3) makes the cold-flow origin of thick discs a viable hypothesis, in combination with its ability to explain the age and chemical duality seen in MW[10].

Galactic bulges, including that of MW, are complex structures[23], and their origin(s) is not yet clear[24]. One possibility is that the gas-rich discs in young galaxies develop star-forming dense clumps due to gravitational instability, which subsequently sink toward the galactic center due to dynamical friction to form a bulge[25]. Alternatively, clumps may have been already formed in the accreting filaments of cold gas[26]. This scenario relies on the longevity of the clumps, which are constantly subject to disruption by feedbacks from star formation inside. The hybrid nature of the halo gas in Domain C may help those clumps survive longer owing to high external gas pressure exerted by the surrounding hot gas component. Interestingly, the essentially same mechanism has been proposed[27] as an origin of low-mass elliptical galaxies in galaxy clusters, where the pressure confinement by hot intra-cluster gas is highly plausible. Galactic bulges have similar sizes, masses, and densities to those compact galaxies[28], further supporting this pressure-confined bulge formation hypothesis.

Present study predicts a large structural variety of high-redshift galaxies. For the redshift range, 2<z<3, we will witness three kinds of evolutional status (Fig.1). Galaxies with small masses will be in Domain A and making thick discs. Intermediate-mass galaxies, staying in Domain C, will be forming bulges in addition to already-formed thick discs. Finally, the most massive galaxies will have already entered Domain B, creating thin discs. Present-day observations can reach such high redshifts, providing interesting new insights. For example, star formation activity in galaxies at $2 < z$ exhibits mass-dependent characteristics, with more massive galaxies showing stronger quenching (i.e., cessation of star formation) in central parts, hinting at the existence of matured bulges[29]. This seems to agree partly with the above prediction although some of massive galaxies may evolve into today's elliptical galaxies by mergers. Other studies show that a large portion of high-redshift galaxies are discy and clumpy[30], which may be in the thick-disc forming phase before bulge and/or thin disc formation. Future observational and theoretical advances are expected to reveal the diversification process of galaxies in the early Universe more than 10 Gyr ago.

## Methods

### The disc galaxy evolution model

In the present work, evolution of a disc galaxy is calculated in two steps.

In the first step, the gas accretion rate at each time step is calculated in the same manner as described in the previous paper[31]. Following the result of cosmological numerical simulations[32], the growth of the virial mass of the dark matter halo with redshift $z$ is specified as $M_{\text{vir}}(z) = M_{\text{vir},0}\, e^{-2Kz/C}$, where $M_{\text{vir},0}$ is the present virial mass, $K$ is a constant 3.7[33], and the halo concentration parameter $C$ at the present epoch is given by
$$\log C = 0.971 - 0.094 \log (M_{\text{vir},0} / 10^{12}\, h^{-1} M_\odot)$$
The primordial gas gathered by the dark matter halo is assumed to have the same density profile as the halo, namely the Navarro-Frenk-White profile, with the adopted cosmic baryon fraction of 0.17. The gas accretion rate in the shock-heating regime (Domain B: $M_{\text{vir}} > M_{\text{shock}}$ and $M_{\text{vir}} > M_{\text{steam}}$) is calculated by assuming that the gas is heated to the virial temperature when entering the host halo and accretes to the disc plane with the timescale of radiative cooling (or with the freefall time if the cooling time is shorter). The accretion rate in the cold-flow regime (Domain A: $M_{\text{vir}} < M_{\text{shock}}$) is calculated assuming that the gas accretes with the freefall time. In Domain C for which $M_{shock} < M_{vir} < M_{stream}$ is satisfied, the halo gas has a 'hybrid' structure in which cold gas flows in through the surrounding hot gas. We assume that all the gas during this period accretes as cold flows. Allotting a certain fraction of the gas to hot-mode accretion does not change the results significantly if it does not exceed about 50%. Modification from the previous treatment[10,31] is as follows. First, the radiative cooling time was calculated assuming the metallicity of the gas entering the halo of $0.01 Z_\odot$ instead of $0.1 Z_\odot$[34]. Second, corresponding to this low metallicity, the values for $M_{shock}$ and $M_{stream}$ given in Ocvirk et al (2008)[17] are used instead of the recipe given in Dekel & Birnboim (2006)[9]. These changes do not affect the conclusions.

In the second step, evolution of the baryonic components is calculated. Instead of resolving the galaxy into a series of concentric annuli as done in the previous studies[10,31], we neglect its spatial structure here. The gas accretion rate calculated as a function of radius in the first step is integrated over the entire region to give the total accretion rate at each time. A galaxy is assumed to be composed of three components: a thin disc, a thick disc, and a bulge. Masses of these respective components at each time characterize the evolutional status of the galaxy.

The mass growth of each component is calculated, assuming that the star formation rate is proportional to the accretion rate. This is a reasonable assumption justified by the previous result of more detailed treatments[10,31] of star formation process in the disk galaxy evolution model. The total mass of the stars formed is adjusted to the observed stellar mass at present using the relationship between the halo mass and stellar mass for late-type galaxies (Extended Fig.1, **a**). Not all the gas accreted is used to form stars in real galaxies. Indeed, the total gas mass which is calculated to accrete during the entire time in the model is larger than the total stellar mass observed. In actual galaxies, a significant portion is considered to escape by the stellar (supernova) feedback[35] or the action of active galactic nuclei[36]. Detailed mechanisms and efficiency of gas removal in these feedback processes are much debated. Instead of tackling this problem, we choose to study how each mass component of the galaxy is formed under the constraint that the total mass of produced stars is consistent with the observed values.

The growth of each component is completely determined in the above procedure. It should be noted that the accretion rate of the halo gas is used only to calculate the growth of stellar masses. In order to follow the chemical element abundances in the interstellar gas and stars, we must specify the mass of cold interstellar gas at each moment and introduce metal enrichment by supernovae. Mass of the cold gas is assumed to follow the observationally inferred values for a redshift range 0<z<2 [37]. Specifically, the cold gas fraction for the stellar mass $M_{star} = 10^{10.25}, 10^{10.75}$, and $10^{11.4}$ $M_\odot$ is set to 0.11, 0.09, and 0.05 at z=0, and 0.42, 0.32 and 0.22 at z=2, respectively (Extended Fig.1, **b**). The gas fraction at each time step was interpolated (or extrapolated if necessary) from these reference values using the current redshift and stellar mass of the model. Lower and upper cutoff of the gas fraction is imposed at 0.05 and 0.9 respectively. The adopted gas fraction is that for the molecular gas. In addition to the molecular gas, atomic hydrogens often contribute much to the total interstellar gas component in spiral galaxies. However, the atomic gas is mostly distributed in the outer part of the galactic disc where star formation is heavily suppressed[38]. Because molecular hydrogens are considered to be the primary raw material for star formation[39], we consider only the molecular component and assume that it occupies the same region as the star forming disc.

Chemical enrichment due to Type II and Ia supernova (SN) is treated in the same way as in the previous study[10] except slight modifications of stellar yields and Type II SN lifetime. Namely, α-elements of 3.18 M☉ and iron of 0.094 M☉ are added to the interstellar gas for one Type II SN. One Type Ia SN is assumed to return 1.38 M☉ of material to the interstellar gas, of which α-elements and iron occupy 0.438 M☉ and 0.74 M☉, respectively[40]. Type II SNe are assumed to explode not instantaneously as done in the previous works[10,31], but $10^7$ years after the birth of progenitor stars. This lifetime corresponds to about 2 time steps and does not bring about any significant change in the model results. Time step with $7 \times 10^6$ years was used to evolve the model, starting at z=37 (age of the universe of $8 \times 10^7$ years).

Of the sample galaxies plotted in Figure 3, those with bulges are dominated by classical bulge or unresolved central mass components. We combined these components into 'bulges' for comparison with the model because both components are consistent with the scenario that they were formed by the migration of clumps to the galactic center. A small number of galaxies with pseudo-bulges were excluded but including them does not make any significant changes. Boxy or peanut-shaped bulges were excluded because they are considered to be bar structures seen edge-on and therefore of disc origin. Yoachim & Dalcanton (2006)[4] have also investigated the thick-to-thin disc mass ratio, but their ratios are systematically smaller than those plotted in Figure 3 albeit a similar decrease with the galaxy mass. This could be because their sample, comprising edge-on thin galaxies, is biased against thick disc components. We did not use their data. In plotting age and metallicity data[14] in Fig.4, we converted the galaxy circular velocity to the stellar mass using the Tully-Fisher relation[41].

**EXTENDED DATA**

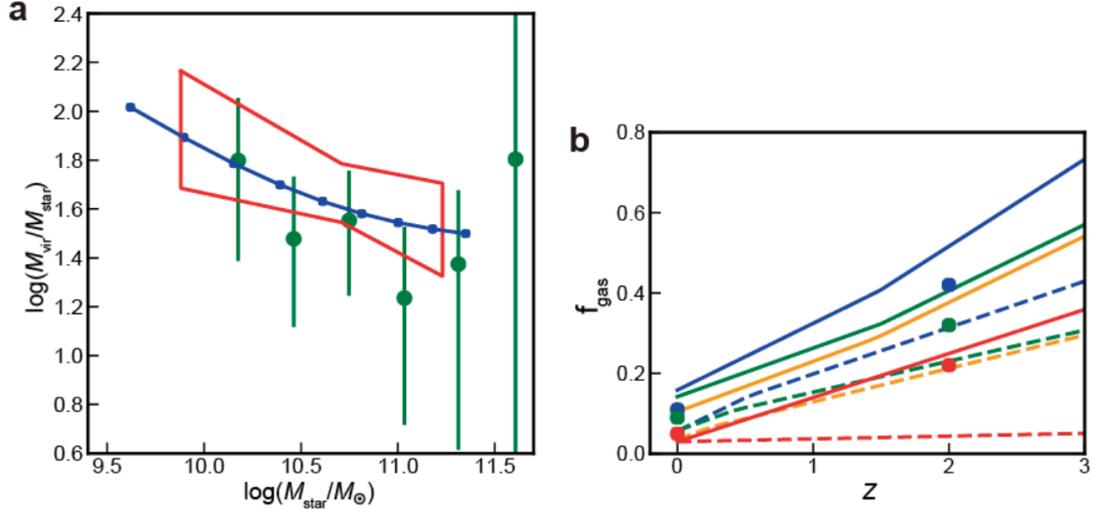

**Extended Data Figure 1. Scaling laws used to construct the model. a**, The ratio of the stellar mass to the virial mass of model galaxies at the present epoch are indicated by blue circles connected by the curve. Observed estimates for disc galaxies from weak gravitational lensing measurements[42] are plotted as green circles with 95% confidence interval. The red line demarcates the distribution area of blue galaxies based on satellite galaxy kinematics[43]. **b**, Mass fraction of the cold interstellar gas to the total baryon mass (i.e., the sum of cold interstellar gas and stars) for four mass bins are indicated by lines (blue: $10.0 < \log M_{star} < 10.5$, green: $10.5 < \log M_{star} < 11.0$, orange: $11.0 < \log M_{star} < 11.4$, and red: $11.4 < \log M_{star}$). Solid and dashed lines indicate the estimated range for each mass bin based on the observation[37]. Circles indicate the reference points used to specify the mass fraction of the interstellar gas in the model for specific values of the stellar mass and redshift by interpolation (or extrapolation) (blue: $\log M_{star} = 10.25$, green: $\log M_{star} = 10.75$, and red: $\log M_{star} = 11.4$).

**Acknowledgments** We acknowledge A. Faisst for providing data for the mass fraction of gas for star forming galaxies.

**Author Information** The author declares no competing financial interest. Correspondence and requests for materials should be addressed to M.N. (noguchi@astr.tohoku.ac.jp).

**Data availability** The data that support the findings of this study are available from the corresponding author upon reasonable request.

**Code availability.** We have opted not to make available the code used to calculate the evolution of the disc galaxy models because it is a part of the integrated program currently in use for other projects.